\documentclass[11pt]{article}
\usepackage{graphicx} \usepackage{dcolumn}   
\usepackage{bm}        \usepackage{amssymb}
\usepackage{float}
\usepackage{amsmath}

\usepackage{hyperref}

\begin{document}

{\centering

{\bfseries\Large Symmetry effects on spin switching of adatoms\bigskip}

C. H\"ubner\textsuperscript{1} , B. Baxevanis\textsuperscript{1,2} , A. A. Khajetoorians\textsuperscript{3,4} and D. Pfannkuche\textsuperscript{1} \\
  {\itshape
\textsuperscript{1}I. Institute for Theoretical Physics, Hamburg University, Germany \\
\textsuperscript{2}Lorentz Institute, Leiden University, Netherlands \\
\textsuperscript{3}Institute of Applied Physics, Hamburg University, Germany \\
\textsuperscript{4}Institute of Applied Physics, Radboud University Nijmegen, Netherlands \\}
(Dated: \today)\\
  }

\begin{abstract}
Highly symmetric magnetic environments have been suggested to stabilize the magnetic information stored in magnetic adatoms on a surface. Utilized as memory devices such systems are subjected to electron tunneling and external magnetic fields. We analyze theoretically how such perturbations affect the switching probability of a single quantum spin for two characteristic symmetries encountered in recent experiments and suggest a third one that exhibits robust protection against surface induced spin flips. Further we illuminate how the switching of an adatom spin exhibits characteristic behavior with respect to low energy excitations from which the symmetry of the system can be inferred. 
\end{abstract}

Recently, single magnetic atoms on surfaces, or so-called magnetic adatoms, have gained a lot of interest for spin-based information storage and processing~\cite{Khajetoorians2011,Loth2012,Miyamachi2013}.  These concepts are mostly based on strong magnetic anisotropy energy~\cite{Hirjibehedin2007,Rau2014}, which reduces spin degeneracy at zero magnetic field, thereby defining preferential spatial orientations of the spin.  While magnetic anisotropy introduces an energy cost for magnetization reversal, countless studies have illustrated that in the presence of strong magnetic anisotropy, individual magnetic adatoms still exhibit rather short lifetimes~\cite{Meier2008,Khajetoorians2011a,Khajetoorians2012} owing to the interplay of the hybridization of the moment bearing orbitals and the underlying substrate. Such observations question the role of both tunneling electrons as well as substrate electrons in dynamical processes of the atomic spin~\cite{Romeike2006,Lorente2009,Delgado2010,Chudnovskiy2014,Bode2012,Rau2014}.  In order to enhance the dynamic stability of such adatoms, strong magnetic coupling between individual spins can be utilized to protect the total spin from fluctuations~\cite{Loth2012,Khajetoorians2013}.  

A different approach, that stabilizes a single magnetic moment of an adatom, was utilized by a particular choice of both spin and underlying substrate symmetry~\cite{Miyamachi2013}. In particular for a three-fold symmetric system with the net magnetic moment $S=8$ of a holmium adatom, it is possible to protect the spin, in the absence of perturbations, from single electron induced spin reversal. It is yet unknown how perturbations, like current-based read out or static magnetic fields which break the symmetry of the system, effect the symmetry caused stability of the spin in such quantum systems.

Here we approach this question with a focus on the prospects of using a specific symmetry to protect a spin from switching. We extract the effective switching rate between the high-spin ground states via all possible spin paths, as experimentally manifested in two-state telegraph noise, utilizing a master equation approach. With comparative analysis we show how a single spin on two-, three- and four-fold symmetric substrates~\cite{Gatteschi2001} responds to temperature, external magnetic field, as well as inelastic tunneling electrons. Higher symmetries are not part of our discussion since they are difficult to realize experimentally. 
We find that the three- and four-fold symmetric systems are both protected against single electron induced ground state switching. Since this protection relies on time-reversal symmetry in the three-fold symmetric system it is highly sensitive to external magnetic fields. In the four-fold case the protection even holds for broken time-reversal symmetry and thus makes it more robust against magnetic fields. On the other hand both systems respond clearly to changes in the energy of tunneling electrons since low energy paths for spin switching between the ground states become accessible. Although spin-flip processes are always possible for the two-fold symmetric system, its dependence on external perturbations is much weaker.  

Such different behavior can be attributed to the symmetry dependent interaction with the underlying crystal field. By expanding the crystal field in terms of spherical harmonics, the interaction with the net spin $\hat{\mathbf{S}}$ of magnetic atoms or clusters can be expressed by various power of spin operators, the so-called Stevens operators~\cite{Stevens1952}. To leading order, the spin Hamiltonian ($\hat{H}_\chi$) can be described by a uniaxial anisotropy term proportional to $\hat{S}_z^2$ and multiaxial terms which are proportional to powers of the raising/lowering operators, $\hat{S}_{+/-}$. In the following the index $\chi \in \{ \mathit{2}, \mathit{3}, \mathit{4} \}$ is used to label the two-, three- and four-fold symmetric Hamiltonian respectively, and we use $\Delta_{01}$ to denote the first excitation energy at zero magnetic field for each symmetry.

\begin{figure}[h!]
  \centering
    \includegraphics[width=\linewidth]{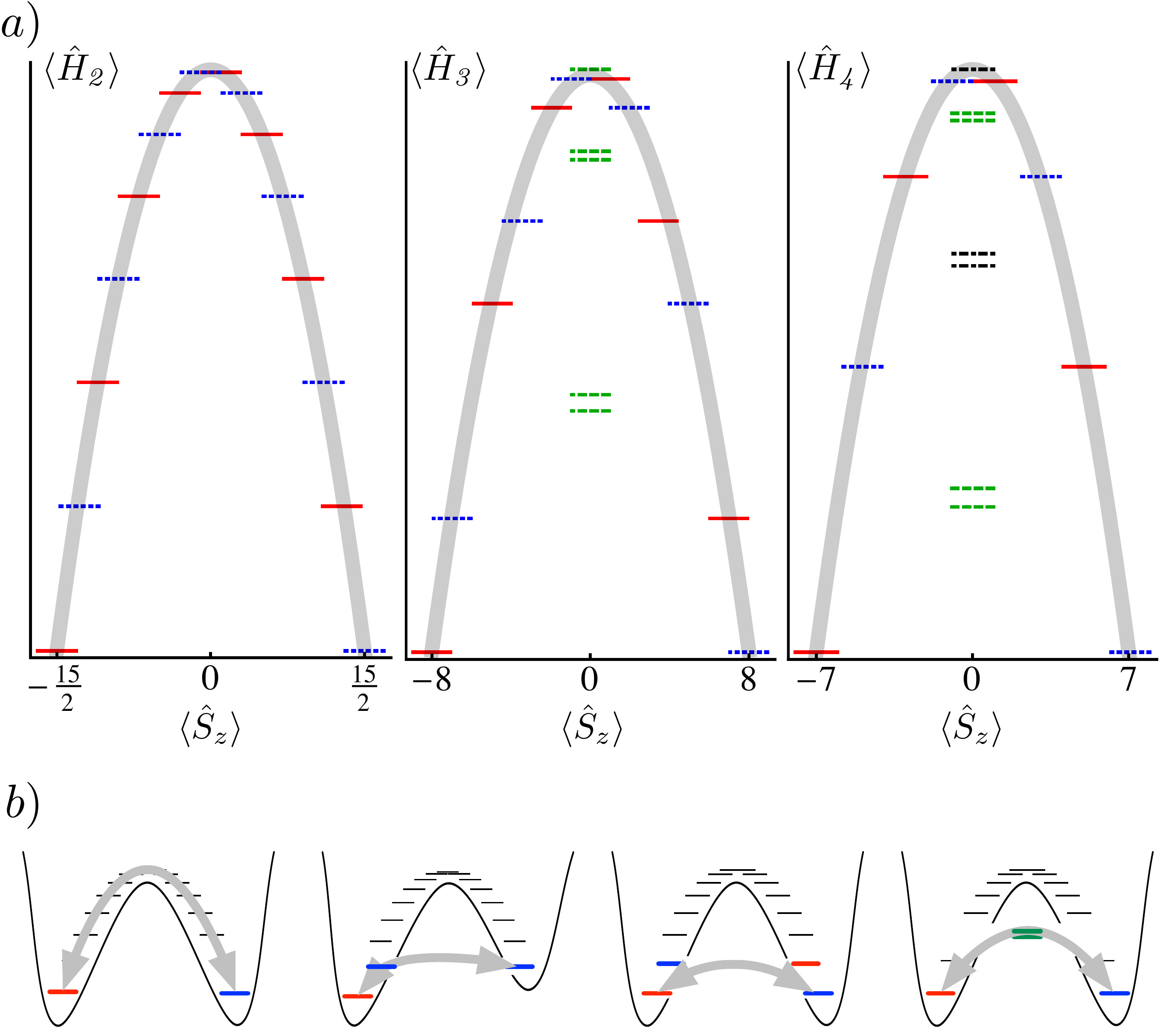}
    \caption{(color online). Energy levels of $\hat{H}_\mathit{2}$ (a left), $\hat{H}_\mathit{3}$ (a center) and $\hat{H}_\mathit{4}$ (a right) as a function of the expectation value $\langle \psi_{\chi,i}^s \vert \hat{S}_z \vert \psi_{\chi,i}^s \rangle$. The parabola indicates the anisotropy barrier. The red, green, blue and black lines indicate $s=-$, $s=+$, $s=0$ and $s=1$ respectively. (b) The different possible paths for spin reversal are illustrated, namely from left to right electron induced switching or ladder transitions over the barrier, quantum tunneling of magnetization, ground state switching, and shortcut tunneling.}
    \label{AnisotropyStates}
\end{figure}

For the two-fold symmetry we can write

\begin{equation}
\hat{H}_\mathit{2} = D_\mathit{2} \hat{S}_z^2 + \tilde{B} \hat{S}_z + E_\mathit{2} (\hat{S}^2_+ + \hat{S}^2_-) 
\end{equation}

where $D_\mathit{2}$ is the uniaxial anisotropy, $\tilde{B}=\mu_{B} g B / \hbar$ the Zeeman energy, $E_\mathit{2}$  the biaxial or transversal anisotropy and the total spin $S_\mathit{2}$. The biaxial anisotropy term is the lowest order contribution leading to mixing of $\hat{S}_z$ eigenstates. Eigenstates $\vert \psi^s_{\mathit{2},i} \rangle$ of $\hat{H}_\mathit{2}$ can be separated in two subgroups $s \in \{+,-\}$, where $i$ is the label within these groups. 

\begin{figure}[h!]
  \centering
    \includegraphics[width=\linewidth]{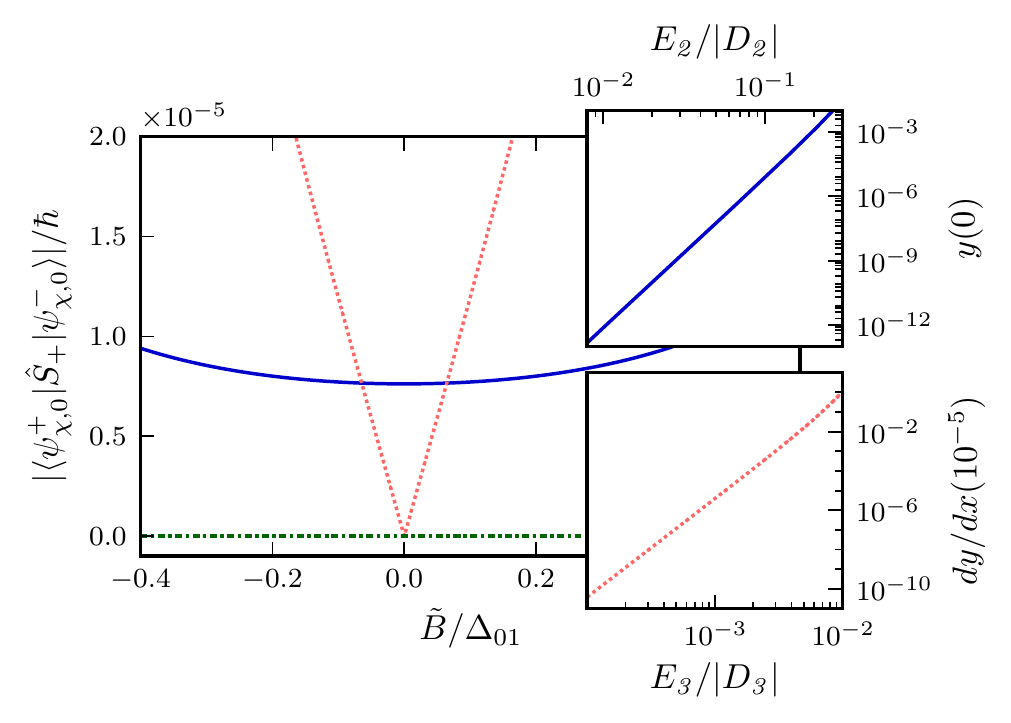}
    \caption{$y (x) := \vert \langle \psi^+_{\chi,0} \vert \hat{S}_+ \vert  \psi^-_{\chi,0} \rangle \vert/\hbar$, as a function of magnetic field $x := \tilde{B}/\Delta_{01}$, where $E_\mathit{2}/|D_\mathit{2}| = 0.1$ is colored solid blue, $E_\mathit{3}/|D_\mathit{3}| = 0.002$ dotted red and arbitrary $E_\mathit{3}/|D_\mathit{3}|$ dash-dotted green. (upper inset) displays the offset of $y$ at $x = 0$ as a function of the biaxial anisotropy $E_\mathit{2}$. (lower inset) shows the linear slope of $y$ at $x=10^{-5}$ as a function of the hexaxial anisotropy $E_\mathit{3}$.}
    \label{Overlap}
\end{figure}

For a half integer spin they are a linear combination of $\hat{S}_z$ eigenstates 

\begin{eqnarray}
\vert \psi^+_{\mathit{2},i}\rangle &=& \sum\limits_{n=0}^{\lfloor S_\mathit{2}\rfloor} c^+_{i,n} \vert S_\mathit{2} - 2 n \rangle \\
\vert \psi^-_{\mathit{2},i}\rangle &=&  \sum\limits_{n=0}^{\lfloor S_\mathit{2}\rfloor} c^-_{i,n} \vert -S_\mathit{2} + 2 n \rangle
\end{eqnarray}

with real coefficients $c^\pm_{i,n}$. We choose a half integer spin number $S_\mathit{2} = 15/2$ for the two-fold symmetric system motivated by a Fe cluster on Cu(111) substrate~\cite{Khajetoorians2013}. For $D_\mathit{2} < 0$ all eigenvalues align along an inverted parabola with respect to the associated expectation value of $\langle \hat{S}_z \rangle$ as depicted in Fig.~\ref{AnisotropyStates}(a) where the two subgroups of eigenstates are explicitly differentiated by color. Due to Kramer’s theorem~\cite{Kramers1930,Klein1952}, direct tunnel coupling between the ground states is forbidden such that $\langle \psi_{\mathit{2},0}^\pm \vert \hat{S}_z \vert\psi_{\mathit{2},0}^\mp \rangle = 0$. Exchange interaction with a single conduction electron leads in lowest order to spin flips which are described by the matrix elements of $\hat{S}_\pm$. Theses so called spectral weights are part of the rates we derive for the master equation used to describe the interaction with the conduction electrons. A single electron can induce a transition between the ground states, since for example the matrix element $\langle \psi^+_{\mathit{2},0} \vert \hat{S}_+ \vert  \psi^-_{\mathit{2},0} \rangle$ gives a non-zero value at zero magnetic field and is proportional to $(E_\mathit{2}/|D_\mathit{2}|)^7 $, as depicted in the upper inset of Fig.~\ref{Overlap}, which can be attributed to the mixing of $\hat{S}_z$ eigenstates by biaxial anisotropy. With increasing magnetic field $|\tilde{B}|$ the probability for the ground state transition induced by a single electron increases further, as shown in Fig.~\ref{Overlap}.

For the three-fold symmetry we can write,

\begin{eqnarray}
\hat{H}_\mathit{3} &=& D_\mathit{3} \hat{S}_z^2 + \tilde{B} \hat{S}_z\\
&+& E_\mathit{3} \left(\hat{S}_z(\hat{S}^3_+ + \hat{S}^3_-) + (\hat{S}^3_+ + \hat{S}^3_-) \hat{S}_z \right). \nonumber
\end{eqnarray}

Here we include only the lowest non-vanishing order of multiaxial anisotropy, which we refer to as the hexaxial anisotropy quantified by its coefficient $E_\mathit{3}$. The eigenstates $\vert \psi_{\mathit{3},i}^s \rangle$ divide in three subgroups $s \in \{+,-,0\}$ for an integer spin and can also be expanded in $\hat{S}_z$ eigenstates. 

\begin{eqnarray}
\vert \psi^+_{\mathit{3},i} \rangle &=& \sum\limits_{n=0}^{\lfloor 2 S_\mathit{3}/3 \rfloor}  c^+_{i,n} \vert S_\mathit{3} - 3 n \rangle \\
\vert \psi^-_{\mathit{3},i} \rangle &=& \sum\limits_{n=0}^{\lfloor2 S_\mathit{3}/3 \rfloor}   c^-_{i,n} \vert - S_\mathit{3} + 3 n \rangle \\
\vert\psi^0_{\mathit{3},i} \rangle &=& \sum\limits_{n=\lceil-S_\mathit{3}/3\rceil}^{\lfloor S_\mathit{3}/3 \rfloor}  c^0_{i,n} \vert 3 n \rangle
\end{eqnarray}

The states are shown color coded in Fig.~\ref{AnisotropyStates}(a) for $D_\mathit{3} < 0$. Unlike the two-fold case, there is a class of eigenstates $s = 0$ which form "within" the potential barrier meaning $\langle \psi_{\mathit{3},i}^0 \vert \hat{S}_z \vert \psi_{\mathit{3},i}^0 \rangle = 0$ and show tunnel splitting at zero magnetic field. With a spin number $S_\mathit{3}=8$, as motivated by the Ho adatom on Pt(111)~\cite{Miyamachi2013}, direct tunneling between the ground states is avoided. In contrast to the two-fold symmetric system, single electron induced tunneling between ground states is forbidden if the spin is not an integer multiple of $3$. Without breaking time-reversal symmetry the matrix elements $\langle \psi^\mp_{\mathit{3},0} \vert \hat{S}_+ \vert  \psi^\pm_{\mathit{3},0} \rangle = \langle \psi^\mp_{\mathit{3},0} \vert \hat{S}_- \vert  \psi^\pm_{\mathit{3},0} \rangle$ vanish~\cite{Miyamachi2013}. As a result, the symmetry of the system protects a given ground state spin from reversal due to single electron fluctuations which is the distinguishing feature of $\hat{H}_\mathit{3}$ as compared to $\hat{H}_\mathit{2}$. To investigate the stability of this symmetry related protection, we apply a ubiquitous magnetic field which breaks the crystal field symmetry. Fig.~\ref{Overlap} shows the increasing probability for switching between the ground states with a single electron with respect to the magnetic field, even in the presence of a small field. The linear increase of switching probability for small magnetic fields is depicted in the lower inset of Fig.~\ref{Overlap} and defines the lower boundary for the switching probability at finite magnetic field. As compared to the quadratic dependence of the biaxial case, this linear behavior witnesses a much stronger sensitivity and a breakdown of the symmetry protection. This needs to be considered if a stray field or neighboring magnetic atoms are present in a spintronic device. 

To lowest non-vanishing order the four-fold symmetric system results in 

\begin{equation}
\hat{H}_\mathit{4} = D_\mathit{4} \hat{S}_z^2 + \tilde{B} \hat{S}_z + E_\mathit{4} (\hat{S}^4_+ + \hat{S}^4_-) 
\end{equation}

with $E_\mathit{4}$ being the coefficient of the multiaxial anisotropy. If the spin is larger than $1$ the eigenstates $\vert \psi^s_{\mathit{4},i} \rangle$ can be arranged in four subgroups $s\in\{+,-,0,1\}$. In this case and an odd integer spin the eigenstates are 

\begin{eqnarray}
\vert \psi^+_{\mathit{4},i} \rangle &=&  \sum\limits_{n=0}^{\lfloor S_\mathit{4}/2 \rfloor}  c^+_{i,n} \vert S_\mathit{4} - 4 n \rangle \label{lground} \\
\vert \psi^-_{\mathit{4},i} \rangle &=&  \sum\limits_{n=0}^{\lfloor S_\mathit{4}/2 \rfloor}  c^+_{i,n} \vert - S_\mathit{4} + 4 n \rangle 
\label{rground}\\
\vert \psi^0_{\mathit{4},i} \rangle &=&  \sum\limits_{n=0}^{\lfloor (2 S_\mathit{4}-1)/4 \rfloor}  c^0_{i,n} \vert S_\mathit{4} - 1 - 4 n \rangle \\
\vert \psi^1_{\mathit{4},i} \rangle &=&  \sum\limits_{n=0}^{\lfloor (2 S_\mathit{4} - 3 )/4 \rfloor}  c^1_{i,n} \vert S_\mathit{4} - 3 - 4 n \rangle.
\end{eqnarray}

For spin $S_\mathit{4} = 1$ subgroup $\vert \psi^1_{\mathit{4},i} \rangle$ would not exist. The eigenstates labeled with $s=0$ and $1$ are similar to the ones "within" the barrier of the three-fold symmetry. Spin switching in a four-fold symmetric system has been studied experimentally by placing a Co atom on the oxygen site of a MgO(100) substrate \cite{Rau2014}. For a spin of $S=3/2$ we do not expect protection of the equilibrium state by symmetry and therefore focus on odd integer spin systems. Usually, the effective spin of an adatom can not be inferred from its free magnetic moment due to surface hybridization and screening of the magnetic moment. Without referring to a specific experimental setup we choose a spin of $S_\mathit{4}=7$ for the following considerations. First it should be of comparable size with the other systems, second it needs to be an integer spin in order to protect it from induced ground state switching with a single electron and third it must be an odd integer to have the ground states belonging to different subgroups. Similar to the three-fold, the four-fold symmetric system is protected against single electron induced ground state switching since $\langle \psi^\mp_{\mathit{4},0} \vert \hat{S}_+ \vert  \psi^\pm_{\mathit{4},0} \rangle = \langle \psi^\mp_{\mathit{4},0} \vert \hat{S}_- \vert  \psi^\pm_{\mathit{4},0} \rangle$ vanishes for arbitrary magnetic field and is shown in Fig.~\ref{Overlap}. It can be shown analytically from equation ~\ref{rground} and ~\ref{lground} or by looking at the subgroups in Fig.~\ref{AnisotropyStates} that at least two coherent electron processes with $\hat{S}^2_\pm$ are needed to cause induced tunneling between the ground states. These processes are highly improbable in the case of weak tunnel coupling which is considered during the following. 

The reversal of the spin between the two ground states is induced by elastic and inelastic spin flips with conduction electrons. Having a scanning tunneling microscope in mind, the conduction electrons are generated from a spin polarized tip and a non-magnetic substrate. They interact with the spin $\hat{\mathbf{S}}$ via exchange interaction described by an Appelbaum Hamiltonian~ \cite{Appelbaum1967}

\begin{equation}
\hat{H}_t = \frac{1}{2}\sum_{rr'kk'\sigma\sigma'}v_r v_{r'} a^{\dagger}_{rk\sigma} {\vec{\sigma}}_{\sigma,\sigma'}\cdot \hat{\mathbf{S}} a_{r'k'\sigma'},
\label{Appelbaum}
\end{equation}

with the annihilation (creation) operators $a_{rk\sigma}^{(\dagger)}$ in tip $r=T$ and substrate $r=S$ and the vector of Pauli matrices  $\vec{\sigma}_{\sigma,\sigma'}$ associated with the spin of the tunneling electrons with momentum $k$. In a perturbative expansion up to fourth order in the coupling $v_r$ and tracing over the electron reservoir degrees of freedom \cite{Delgado2010} a master equation

\begin{equation}
\frac{dP_{\alpha}}{dt}=\sum_\beta \left(\mathcal{W}_{\alpha \beta} P_\beta - \mathcal{W}_{\beta \alpha} P_\alpha \right)
\label{master-eqn.}
\end{equation} 

for the reduced density matrix can be derived. It describing changes in the occupation probability of $\hat{H}_\chi$ eigenstates. Within the rates 

\begin{equation}
\mathcal{W}_{\alpha\beta}=\pi \sum_{rr'\in \{ \text{tip,sub} \} } |v_r v_{r'} |^2 \Sigma^{r r'}_{\alpha\beta} \, \zeta\left(\mu_r - \mu_r' - \Delta_{\alpha\beta} \right) 
\end{equation} 

including the spectral weight $\Sigma^{r r'}_{\alpha\beta}$ and the energy selection rule $\zeta(x )= \frac{x}{1-\exp(-\frac{x }{k_b T})}$. The argument of $\zeta$ includes the chemical potential $\mu_r$ associated with an electron reservoir (tip or substrate) and the energy difference $\Delta_{\alpha\beta}=\epsilon_\alpha - \epsilon_\beta$ between eigenstates $\vert \alpha \rangle$ and $\vert \beta \rangle$. The spectral weight 

\begin{eqnarray}
\Sigma^{r r'}_{\alpha\beta} &=& |\langle\alpha\vert\hat{S}_+\vert\beta\rangle|^2\rho_{r\downarrow}\rho_{r'\uparrow}+ |\langle\alpha\vert\hat{S}_-\vert\beta\rangle|^2\rho_{r\uparrow}\rho_{r'\downarrow} \\
& &+ |\langle\alpha\vert\hat{S}_z \vert\beta\rangle|^2 \left(\rho_{r\uparrow}\rho_{r'\uparrow}+\rho_{r\downarrow}\rho_{r'\downarrow} \right) \nonumber,
\end{eqnarray}

includes matrix elements of spin operators $\hat{S}_\pm$, standing for a change of the single spins orientation. The spin densities $\rho_{r\sigma}$ describes whether the conduction electron has flipped its spin or remained in the same orientation during tunneling. The reservoir polarization $\mathcal{P}_r=\frac{\rho_{r\uparrow}-\rho_{r\downarrow}}{\rho_{r\uparrow}-\rho_{r\downarrow}}$ is choses to be zero for the unpolarized substrate and $0.1$ for the tip. We consider the case in which renormalization of energy levels from scattering on electrons~\cite{Oberg2013} can be neglected. This means the tunnel coupling has to be sufficiently small compared to the temperature. Under this condition an electron bath does not destroy the coherence between the $\hat{S}_z$ states contributing to the $\vert \psi^{0/1}_{\chi,i} \rangle$ groups. By solving the master equation we identify the dominant switching rate $\Gamma$ between the two polarized spin states $\vert \psi^+_{\chi,0} \rangle$ and $\vert \psi^-_{\chi,0} \rangle$. All possible paths are included in the switching rate and can be characterized as depicted in Fig.~\ref{AnisotropyStates}(b). Each path is effected differently by external perturbations such as temperature, applied voltage or magnetic field. This makes their contribution to the total switching rate $\Gamma$ distinguishable in specific parameter regimes. All rates will be given in units of the total spin flip rate $\Gamma_0 = \pi \hbar^2 v^2_{T} v^2_{S} (\rho_{T\uparrow} \rho_{S\downarrow} + \rho_{T \downarrow} \rho_{S\uparrow})/|D_\chi|(2S_\chi-1)$ for electrons inelastically tunneling from the tip to the surface.

\begin{figure}[h!]
  \centering
    \includegraphics[width=\linewidth]{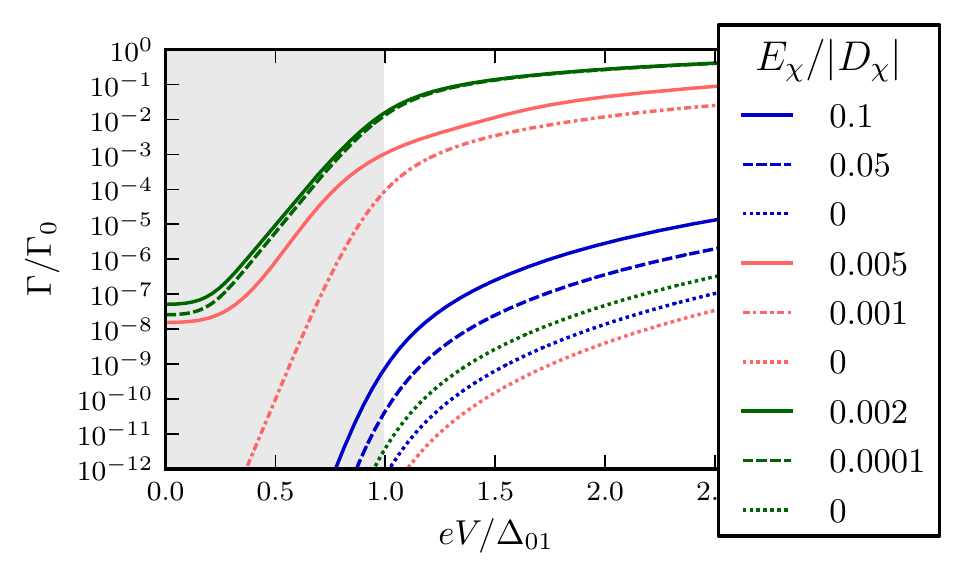}
    \caption{(color online). Switching rate $\Gamma$ between the ground states as a function of applied bias voltage $eV$ between the tip and the surface. The results for $\hat{H}_\mathit{2}$ are colored blue, for $\hat{H}_\mathit{3}$ red and $\hat{H}_\mathit{4}$ green. Multiaxial anisotropy $E_\chi/|D_\chi|$ allows temperature induced excitations below the threshold voltage  $eV = \Delta_{01}$ marked by the gray region. ($\vert v_T \vert/\vert v_T \vert = 0.15$, $\tilde{B}=0$, $k_{\rm{B}}T = 0.05 \Delta_{01}$)}
    \label{SurmountBarrier}
\end{figure}

In all three cases at least $2S_\chi$ sequential spin flip processes are needed to surmount the anisotropy barrier along the spin ladder. Nevertheless, due to strong relaxation generated by substrate electrons, typically more electrons are needed. Moreover, each electron requires a minimum energy of $\Delta_{01}$ for an inelastic scattering event with the adatom spin $S$, which corresponds to the first excitation energy of the spin system. Such electron energies are generated either by the applied voltage $eV$ between the tip and substrate or from temperature $k_{\rm{B}}T$. With just uniaxial anisotropy and with small temperatures $k_bT \ll \Delta_{01}$ a threshold voltage $eV = \Delta_{01}$ denotes the onset of spin switching due to the aforementioned ladder processes as shown in Fig.~\ref{SurmountBarrier} (bright region). The system with the largest spin $S_\mathit{3}=8$ shows the smallest switching rate, since more inelastic excitations are needed to reverse the spin due to the larger number of ladder states. 

At zero voltage and constant $k_bT/\Delta_{01}$ temperature induced transitions between the ground and the first excited state allow switching even below the first excitation energy $eV=\Delta_{01}$ as depicted in Fig.~\ref{SurmountBarrier} (greyed region). The mixing of $\hat{S}_z$ eigenstates with multiaxial anisotropy leads to an increase of the rate between the ground and the first excited state. Additionally the energy levels of the spin states are shifted by the multiaxial anisotropy which has an effect on the rates between excited states. 

Especially for the three- and four-fold symmetric system the excited states $\vert \psi^{0/1}_{\chi,i} \rangle$ become accessible and establish an extra switching path that is absent under two-fold symmetry. For the four-fold symmetry the switching is dominated by a path via the first excited state. Therefore changes of the spectrum or mixing of high energy spin states with $E_\mathit{4}$ has only a small effect on the switching rate. 

%Despite the absolute value of the switching rate $\Gamma_\chi$ depending on $E_\phi/E_\psi$, the characteristic behavior with respect to the voltage is similar in both systems. 

A single electrons that induces elastic quantum tunneling between the ground states mainly originates from the unpolarized substrate since the relative coupling $v_T/v_S \ll 1$ is much stronger. Rates between two degenerate states for an unpolarized substrate become

\begin{equation}
\mathcal{W}_{\psi^\mp_{\chi,i} \psi^\pm_{\chi,i}} \propto \left( |\langle \psi^\mp_{\chi,i} \vert\hat{S}_+\vert \psi^\pm_{\chi,i} \rangle|^2+ |\langle \psi^\mp_{\chi,i} \vert\hat{S}_-\vert \psi^\pm_{\chi,i}\rangle|^2 \right) k_b T
\end{equation}

and represent transitions due to electron-induced quantum tunneling. A single electron can transfer its spin to induce a transition between the ground states since they are a mixture of $\hat{S}_z$ eigenstates. For half integer adatom spin a single electron can always induce transitions between the two highest lying degenerate states. From the rate one can distinguish two situations for which transitions between lower lying degenerate spin states are forbidden: (i) the complete absence of multiaxial anisotropy. (ii) the symmetry of the system prohibits electron induced quantum tunneling as is the case with the three- and four-fold symmetry in combination with a certain choice of spin. $S_\mathit{3}$ being not an integer multiple of three and $S_\mathit{4}$ being an odd integer leads protection caused by symmetry. While this protection is observed in the lifetime in ref.~\cite{Miyamachi2013} for the three-fold symmetric system it has not been studied yet for four-fold symmetry. 

\begin{figure}[h!]
  \centering
    \includegraphics[width=\linewidth]{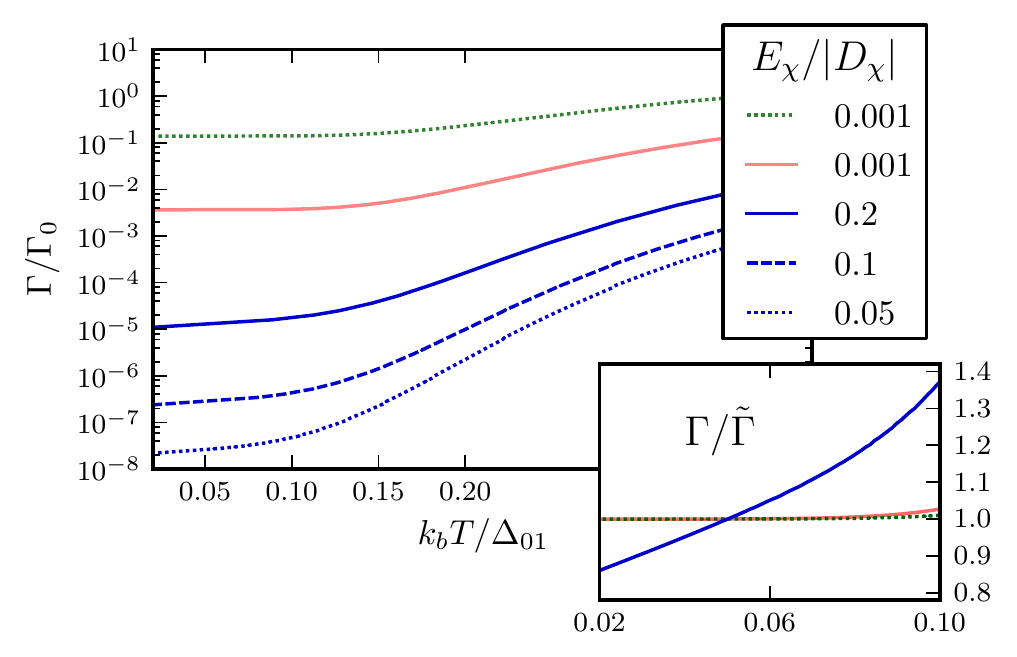}
    \caption{(color online). Temperature dependence of switching rate $\Gamma$ at $eV = 1.5 \Delta_{01}$. The results for $\hat{H}_\mathit{2}$ are colored blue, for $\hat{H}_\mathit{3}$ red and $\hat{H}_\mathit{4}$ green. The inset magnifies the low temperature region and is normalized to $\tilde{\Gamma}=\Gamma(k_bT = 0.05 \Delta_{01})$. ($\vert v_T \vert/\vert v_T \vert = 0.15$, $\tilde{B}=0$) }
    \label{Temperature}
\end{figure}

Fig.~\ref{Temperature} depicts the effect of electron-induced quantum tunneling on the switching rate for all systems as a function of temperature. At $k_bT>0.15 \Delta_{01}$, thermally induced switching over the barrier results in Arrhenius-like behavior. The largest effect of electron-induced quantum tunneling can be seen at temperatures $k_{\rm{B}}T < 0.1 \Delta_{01}$ (inset of Fig.~\ref{Temperature}) in which thermal excitations can be neglected for the chosen multiaxial anisotropy values. The absence of magnetic field results in the highest degree of degeneracy. For $\hat{H}_\mathit{2}$ multiple channels for electron induced quantum tunneling are accessible due to a finite voltage $1.5 eV/\Delta_{01}$. For $\hat{H}_\mathit{3}$ on the other hand electron induced quantum tunneling is forbidden due to symmetry and a plateau appears for temperatures $k_{\rm{B}}T < 0.1 \Delta_{01}$. The same plateau appears for $\hat{H}_\mathit{4}$. This makes the three- and four-fold symmetric system more robust than the two-fold against surface electron-induced switching in the low temperature regime. For three-fold symmetry this statement only holds in the absence of perturbations breaking time-reversal symmetry while the protection by the four-fold symmetry is valid even with magnetic field and can be seen in Fig.~\ref{Overlap} and shown below in Fig.~\ref{Protection}(b).

\begin{figure}[h!]
  \centering
    \includegraphics[width=\linewidth]{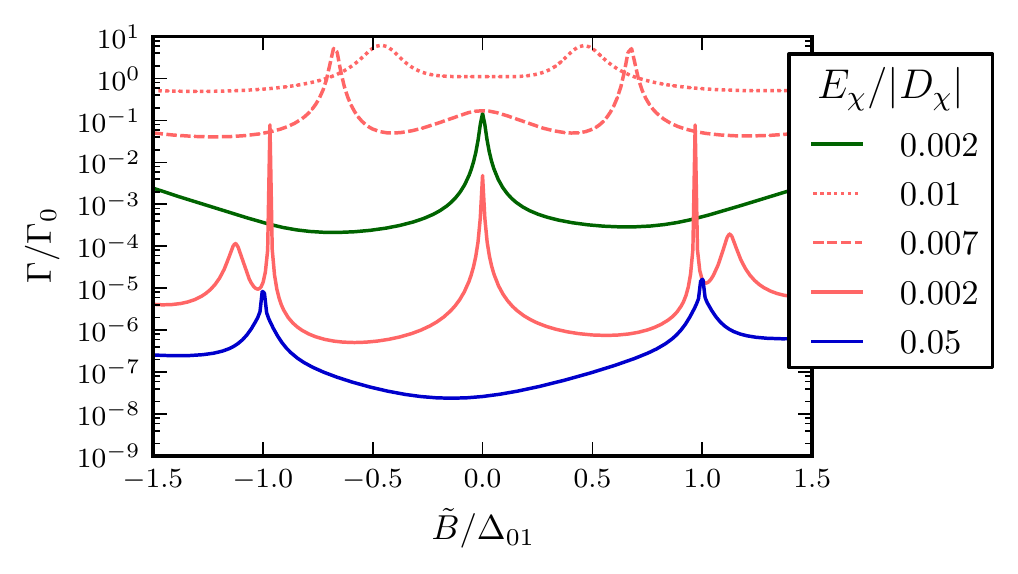}
    \caption{(color online). Switching rate $\Gamma$ in dependence of magnetic field at $eV = 1.5 \Delta_{01}$ and $k_b T = 0.05 \Delta_{01}$. The results for $\hat{H}_\mathit{2}$ are colored blue, for $\hat{H}_\mathit{3}$ red and $\hat{H}_\mathit{4}$ green. The center resonance is a result of short cut tunneling and becomes visible for $\hat{H}_\mathit{3}$ at $E_\mathit{3}/|D_\mathit{3}|<0.01$. The narrow resonance for the three-fold symmetry belongs to quantum tunneling of magnetization and shifts from $|\tilde{B}| / \Delta_{01} \approx 0.45$ to $1.0$ with the change of hexaxial anisotropy. The broad side peak is due to restored short cut tunneling by magnetic field. The tip polarization is 10\%. }
        \label{QTM}
\end{figure}

 At larger magnetic fields, namely $\tilde{B} \approx \Delta_{01}$, relaxation channels via tunnel mixed spin states open resulting from quantum tunneling of magnetization~\cite{Gatteschi2006}. A finite $E_\chi$ is needed in order to obtain this mixing. The stronger the mixing the more robust is the relaxation channel against variation of the magnetic field from the resonance condition. This is manifested in Fig.~\ref{QTM} as peaks in the switching rate. For $E_\mathit{2}/|D_\mathit{2}|=0.05$ eigenstates are only weakly perturbed $\hat{S}_z$ eigenstates, however at $\vert\tilde{B}\vert= \Delta_{01}$ all unperturbed states cross pairwise leading to an efficient mixing in the presence of finite $E_\mathit{2}$. $\hat{H}_\mathit{2}$ has the strongest mixing between spin states at the top of the barrier.

Although for the four-fold symmetric system mixing appears also at the top of the barrier, the resonance condition is shifted to $\vert\tilde{B}\vert= 2 \Delta_{01}$. This is because twice the magnetic field in units of $\Delta_{01}$ is needed for aligning states of the same subgroup and to ensure efficient mixing. The characteristic peak, present at zero magnetic field, can be related to inelastic electron induced short cut tunneling created by states from the subgroup $\vert \psi^{0/1}_{\chi,i} \rangle$ as depicted in Fig.~\ref{AnisotropyStates}(b).
The largest contribution to switching via quantum tunneling of magnetization in $\hat{H}_\mathit{3}$ results from mixing of spin states in the valley of the barrier namely $\vert \psi^\pm_{\mathit{3},0} \rangle$ and $\vert \psi^\mp_{\mathit{3},1} \rangle$, if $\vert \psi^\mp_{\mathit{3},0} \rangle$ is the ground state.  For $E_\mathit{3}/|D_\mathit{3}|=0.01$ a single resonant tunneling channel emerges at $\vert\tilde{B}\vert \approx 0.45 \Delta_{01}$. Since for $E_\mathit{3}/|D_\mathit{3}|=0.01$ the hexaxial anisotropy already shifts the energy levels, the resonance conditions occur at different magnetic fields. This leads to a sequence of single channels that open for relaxation. 
Lowering the anisotropy to $E_\mathit{3}/|D_\mathit{3}| \rightarrow 0.002$ has two effects: (i) the resonance peaks from quantum tunneling of magnetization shift to the resonance condition of the unperturbed system $\vert\tilde{B}\vert= \Delta_{01}$. (ii) an underlying substructure becomes visible with a central resonance peak that can be related to inelastic electron induced short cut tunneling created by states from the subgroup $\vert \psi^0_{\mathit{3},i} \rangle$. The side peak at $\vert\tilde{B}\vert\approx 1.2 \Delta_{01}$ comes from a short cut that is reestablished with magnetic field. 
The features in Fig.~\ref{QTM} can be used to differentiate between the different symmetries. The absence of a central resonance peak makes it possible to distinguish the two-fold from the three- and four-fold symmetric system. The presence of a resonance peak from quantum tunneling of magnetization at $\tilde{B}<\Delta_{01}$ is a indication for a three-fold symmetric system since the peak appears at higher magnetic fields for the four-fold symmetric system.

\begin{figure}[h!]
  \centering
    \includegraphics[width=\linewidth]{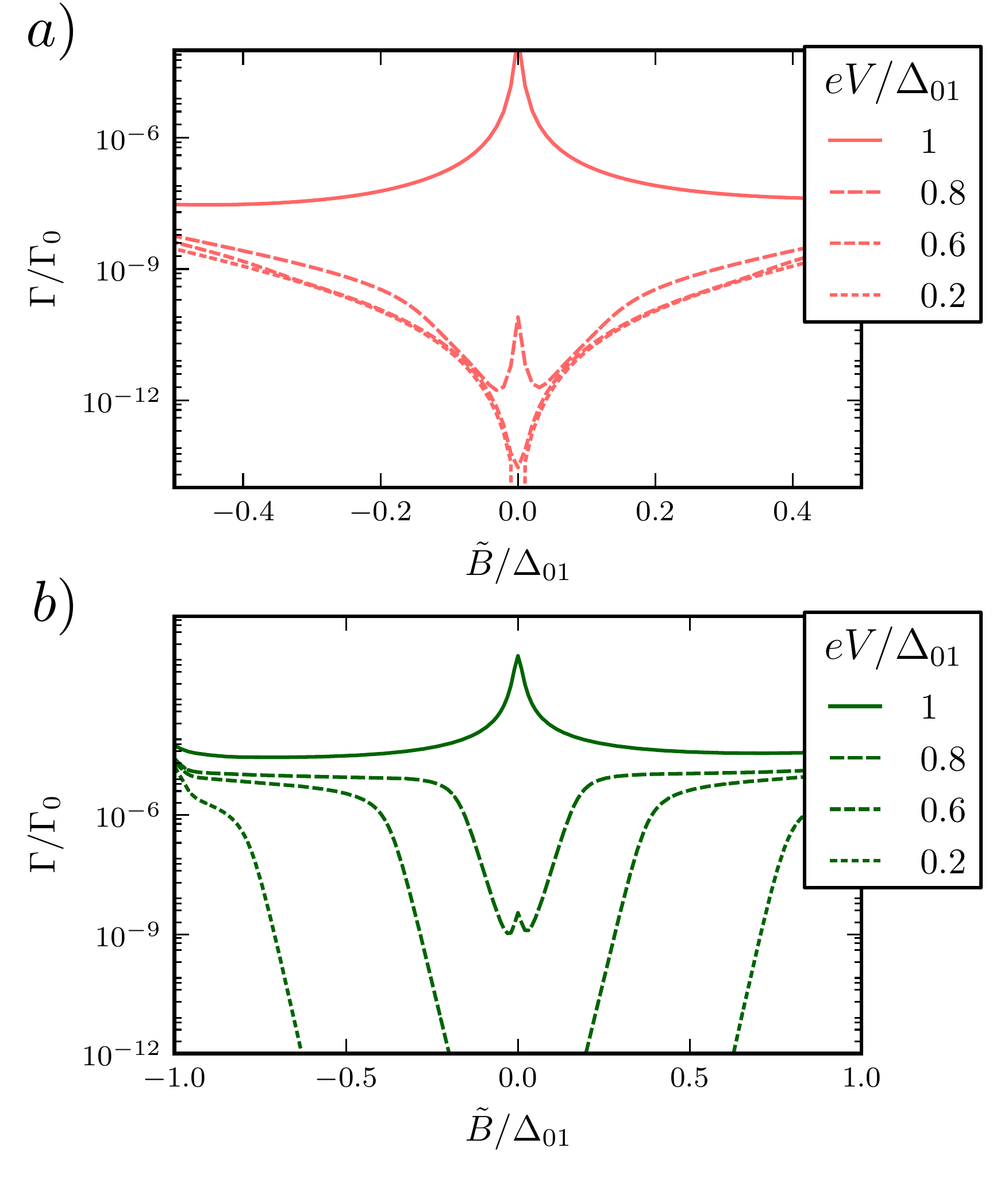}
    \caption{(color online). The switching $\Gamma$ rate at $E_\chi/|D_\chi|= 0.002$ as a function of magnetic field for (a) three-fold and (b) four-fold symmetry. Decreasing voltage $eV/\Delta_{01} \in \{ 1,0.8,0.6,0.2 \} $ is shown in decreasing order of the rate. The tip polarization is 10\% and the temperature is $k_BT=0.01\Delta_{01}$. }
        \label{Protection}
\end{figure}

 Fig.~\ref{Protection} shows the disappearance of the central resonance from short cut tunneling at small voltages. If the electron energy, resulting from voltage and temperature, exceeds the threshold $\Delta_{01}$ inelastic scattering can lead to the first and second excitation and thus switching becomes effective through the shortcuts. Below the threshold voltage and with exponentially suppressed thermal excitations the switching rate goes to zero since the ground state switching is prohibited by symmetry. In the case of $\hat{H}_\mathit{2}$ the rate would not go to zero since electron induced ground state switching is alway present.
For $\hat{H}_\mathit{3}$ a finite magnetic field destroys the time-reversal symmetry and thus an increasing switching rate is observed even though excitations are suppressed. Above the threshold voltage the switching rate $\Gamma$ decreases with magnetic field since the eigenstates $\vert \psi^0_{\mathit{3},i} \rangle$ become well separated on both sides of the anisotropy potential and the short cut is closed. 
Same behavior can be seen for $\hat{H}_\mathit{4}$ if the electrons have enough energy for the first excitation. Below the threshold, when this is exponentially suppressed, the four-fold symmetric system shows a broad range of magnetic fields where the spin is symmetry protected from switching. For a device that utilizes the symmetry of the system to have a stable orientation of the spin we would therefore recommend a four-fold symmetric system with odd integer spin. Still, accompanied with the protection by symmetry is the possibility of switching via low energy paths through the anisotropy barrier. Only if thermal excitations can be suppressed the protection mechanism can be used efficiently. 

In conclusion we compared the qualitative behavior of the switching rate between two polarized states of a single spin in a two-, three- and four-fold symmetric system with respect to external perturbations such as magnetic field, temperature and spin excitations due to conduction electrons. We found that the protection against ground state transitions, induced by a single electron, in the three-fold symmetric system can be destroyed by a magnetic field. The four-fold symmetric system also has a protection against single electron induced ground state transitions which is independent of magnetic field. This makes the switching in the three-fold symmetric system more sensitive to time-reversal symmetry breaking in contrast to the two- and four-fold symmetric systems. Transitions to excited spin states, via inelastically tunneling electrons, make multiple fast paths for switching across the anisotropy barrier accessible. The three- and four-fold symmetry provide rapid switching through short cuts in the barrier that are missing under two-fold symmetry. The characteristic behavior makes it possible to distinguish the symmetry by measuring the switching rate as a function of magnetic field.

We acknowledge funding through SFB925, GrK1286, SFB668 and Emmy Noether KH324/1-1. We want to thank KIT members for invitation and discussion, especially Christian Karlewski, Wulf Wulfhekel and Gerd Sch\"on.

\end{document}